\documentclass[compsoc, conference, a4paper, 10pt, times]{IEEEtran}

\usepackage{cite}
\usepackage{amsmath,amssymb,amsfonts}
\usepackage{algorithmic}
\usepackage{graphicx}
\usepackage{textcomp}
\usepackage{xcolor}
\usepackage{booktabs}
\usepackage[hidelinks]{hyperref}
\usepackage{balance}

\usepackage{url}            
\usepackage{amsfonts}       
\usepackage{nicefrac}       
\usepackage{microtype}      

\usepackage{amsmath}
\usepackage{relsize}
\usepackage{makecell} 

\begin{document}

\title{Your Email Address Holds the Key: Understanding the Connection Between Email and Password Security with Deep Learning}


\author{
\IEEEauthorblockN{Etienne Salimbeni*}
\IEEEauthorblockA{\textit{EPFL} \\
Lausanne, Switzerland}
\and
\IEEEauthorblockN{Nina Mainusch*}
\IEEEauthorblockA{\textit{EPFL} \\
Lausanne, Switzerland}
\and
\IEEEauthorblockN{Dario Pasquini}
\IEEEauthorblockA{SPRING Lab, \textit{EPFL} \\
Lausanne, Switzerland}
}

\maketitle
\def\thefootnote{*}\footnotetext{These authors contributed equally to this work. The order of their names has been determined by flipping a coin.}\def\thefootnote{\arabic{footnote}}
\def\thefootnote{}\footnotetext{Presented at the IEEE Symposium on Security and Privacy workshop for deep learning and security 2023. (DLSP'23)}\def\thefootnote{\arabic{footnote}}

\begin{abstract}
In this work, we investigate the effectiveness of deep-learning-based password guessing models for targeted attacks on human-chosen passwords.
In recent years, service providers have increased the level of security of users' passwords.
This is done by requiring more complex password generation patterns and by using computationally expensive hash functions.
For the attackers this means a reduced number of available guessing attempts, which introduces the necessity to target their guess by exploiting a victim's publicly available information.
In this work, we introduce a context-aware password guessing model that better capture attackers' behavior.
We demonstrate that knowing a victim's email address is already critical in compromising the associated password and provide an in-depth analysis of the relationship between them. 
We also show the potential of such models to identify clusters of users based on their password generation behaviour, which can spot fake profiles and populations more vulnerable to context-aware guesses.
The code is publicly available at \texttt{https://github.com/spring-epfl/DCM\_sp}.





\end{abstract}



\section{Introduction}

User-generated passwords are up until today the main authentication mechanism in modern software applications. 
They are convenient to implement and deploy for developers, and they are straightforward to use for the end-users.
On the downside, humans tend to choose their passwords based on convenience rather than security, enabling password guessing attacks via dictionaries.

Since humans often use repetitive patterns for their passwords, software tools have been developed that generate passwords based on word-lists and manually curated password mangling rules.
This is intended to model human password generation behavior, for instance, the "e" in a word is replaced by a "3", or a number is appended to a word.
Well known tools of this kind are John the Ripper and HashCat \cite{openwall, hashcat}.

Removing the need to manually create the password generation rules, Markov models and probabilistic context-free grammars (PCFGs) statistically extrapolate rules from publicly available password leaks \cite{PCFGs2009}. 
Markov models learn to generate passwords based on preceding context characters and PCFGs learn the underlying pattern probabilities of a rule being used in a password \cite{PCFGs2009}.

Recently, the performance of the aforementioned methods has been surpassed by a new class of models, artificial neural networks.
An intuitive modern network architecture to model a password distribution space is a recurrent neural network (RNN).
RNNs are sequence-to-sequence models with feedback connections and an internal memory state that remember information about previous elements in the sequence.
This architecture has been shown to outperform previous approaches \cite{Melicher2016}.

However, it is known that real-world adversaries make use of users' publicly available data to improve their attacks~\cite{Pal}.
The additional data can leak information about a user's nationality, relatives, pets, or hobbies, all of which provide information about the conditional password space of a user. 

The importance of investigating the relationship between public information and passwords has been reinforced by recent data breaches and leaks, including the password manager LastPass data leak in December 2022, which exposed sensitive information of millions of users.

To adapt models to the new attack behaviour, sequence-to-sequence models emerged. 
In pass2pass, passwords are generated given a list of leaked password from a user ~\cite{Pal}. In the same direction,~\cite{Zhou} use a set of public information of a user such as the email, phone number, birthday, website domain, nickname, and the last/first name as input to their Transformer model, cf. Section~\ref{Related_work}.

Current research focuses on models performing guessing attacks but does not investigate the relation between the model, the public information, and the user password. 
In this paper we will focus on exactly this relationship, using the user email as representative public information.


\subsection{Contributions}

\begin{figure}[t]
    \centering
    \includegraphics[scale=2.0, page=5] {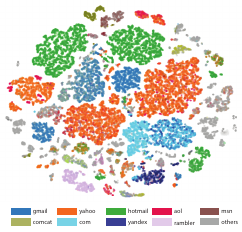}
    \caption{Email subdomain. We show the hidden states of the context-sensitive model and can observe that the model disentangled the structure of the subdomains, which implies that it can extract valuable information about the password from it.
    In gray are domains that are not listed in the legend.}
    \label{fig:subdomain_hidden}
\end{figure}

\begin{figure}[t!]
    \centering
   \includegraphics[scale=2.0, page=6] {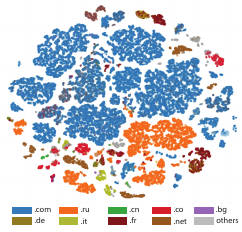}
    \caption{Email topdomain. We show the hidden states of the context-sensitive model decoder and can observe that the model disentangled the structure of the topdomains, which implies that it can extract valuable information about the password from it.
    In gray are domains that are not listed in the legend.}
    \label{fig:topdomain_hidden}
\end{figure}

With our work, we aim to enable a more robust and precise password security evaluation; to inform users about the extent to which they are vulnerable to realistic, adaptive guessing attacks.
Through a comprehensive analysis, we demonstrate the impact of supplementary information on the security of passwords and provide insights on how to enhance password protection.
Our work lays the foundation to develop a password strength control that is not based on generic control factors like length, presence of special symbols, uppercase letters, but displays the real strength of a personalized password.
Our main contributions are:
\begin{itemize}
  \item We investigate the association between email addresses and passwords and demonstrate the feasibility of using context-aware deep learning to model this relationship.
  \item We formally asses the superiority of context-aware attention model over general password generator model using the guess number metric for targeted attacks
  \item We perform a thorough analysis of the change in the latent password space from the base to the context-sensitive model. 
  \item We show that such models expose groups of users with similar password generation behaviour vulnerable to attacks leveraging contextual information, cf. Figure~\ref{fig:subdomain_hidden} and Figure~\ref{fig:topdomain_hidden}.
  
\end{itemize}

\section{Related work}
\label{Related_work}

Password modeling based on neural networks replaced the usage of manually created or statistically inferred password generation rules.
First steps were taken in exploring the effectiveness of simple recurrent neural network architectures for this modeling task ~\cite{Melicher2016}.
Since then, various improvements have been proposed to accurately learn the password distribution space, for example by leveraging the generative adversarial network architecture~\cite{Hitaj, Pasquini2021} or language representation models like BERT~\cite{Li2019} and GPT2~\cite{Biesner2020}.
Those approaches are limited to modeling the unconstrained password space, however, they do not incorporate additional available information.

LSTM networks are a type of recurrent neural network capable of processing elements in sequences. 
They use an internal memory to remember information about previous elements in the sequence \cite{LSTM_EXPLAINED}.

By implementing an LSTM model that maps from the personal information space to the actual password space, Zhou et al. demonstrated that users partially integrate their name, birthday, username, and email prefix into their constructed passwords ~\cite{Zhou}.
Harnessing the fact that intra-user generated passwords can be extremely similar, Pal et al. analyzed the situation where there are previously leaked passwords from a user available ~\cite{Pal}.
They designed an encoder-decoder model called \texttt{pass2pass} and showcased its effectiveness in compromising user accounts.

Employing the Transformer architecture and augmenting it by adding an information weight layer in the data pre-processing, He et al. showed that it surpasses the effectiveness of other models such as a standard Transformer, an encoder-decoder model with attention, and an LSTM, in terms of password guessability ~\cite{He2020}.
They investigated the relationship between personal information such as the email, phone number, birthday, website domain, nickname, and the last- and first name, and the generated password.
Through substring matching, they showed that the email is the most correlated with the password of all those features.

The fact that email addresses are the personal attribute with the best predictive power and that they are frequently available together with a leaked password motivates our research to improve the password space modeling conditioned on a user's email address.
We leverage the LSTM architecture because it has been shown to be effective for generating text in the context of character-level natural language tasks~\cite{Graves13, Sutskever2011GeneratingTW} and is therefore the most promising model to accurately reflect how the latent password distribution changes given the additional information.
%
To investigate the influence of each component of the additional information, we do not just concatenate all the personal information as in He et al.~\cite{He2020}, but we map each feature in a meaningful embedding space first and interpret the individual importance of each feature based on the attention head evaluation score.

\section{Methods}
\subsection{Dataset}
The dataset is comprised of $1.4$ billion clear text credentials.
The credentials originate from the Exploit.in and Anti Public password leaks and contain $252$ breaches. 
We process the data such that each email-password pair has the format: $$\texttt{username@subdomain.topdomain:password}$$

\paragraph{\textbf{Ethical considerations}} We are showing email-password pairs in Section~\ref{section:results} which does not
impair the affected accounts any further, since the dataset
is publicly available.
Although our dataset is publicly available and has been used in other papers \cite{Pal}, all shown plaintext pairs were checked with $haveibeenpwned.com$ to not cause additional harm for the users.

\subsection{Base auto-regressive model}
Our base model is comprised of a two-layer LSTM with $512$ units followed by a dense layer, identical to the model architecture propsed by Melicher et al.~\cite{Melicher2016}. 
See Appendix A.\ref{fig:basemodel} for more details.
The input is mapped into an embedding of size $512$ and passed into the LSTM.
The auto-regressive nature of the model is needed for computing the Monte Carlo estimation discussed in Section~\ref{subsec:guess_numbers}.

\subsection{Context-sensitive attention model}
\begin{figure}[h!]
  \centering
    \includegraphics[scale= 0.8, page=2] {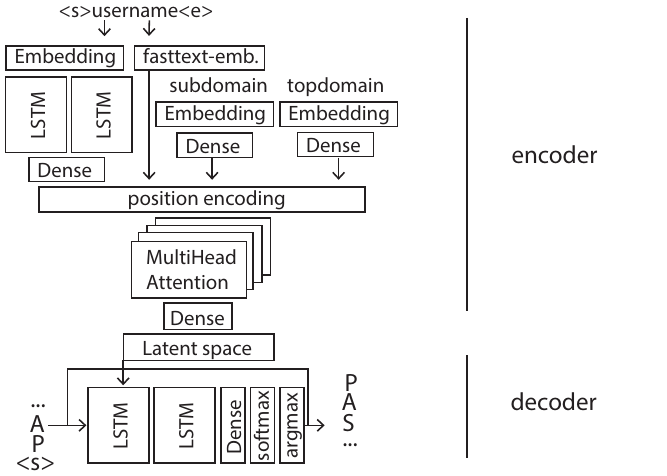}
    \caption{Model architecture of our context-aware model. 
    The embedding of each email segment is reshaped by a dense layer to size $128$ and a position embedding is to them.
    Afterward, it is passed to a $4$-headed attention layer with self-attention.
    The output of the MHA layer is passed to the first LSTM if the decoder has hidden states.  }
    \label{fig:contextmodel}
\end{figure}
The context-sensitive attention model extracts the information from a victim's email address that is critical in compromising the associated password.
Our complete architecture is depicted in Figure~\ref{fig:contextmodel}.
The main components of the attention model are the encoder, decoder, and the self-attention mechanism.

\subsubsection{Encoder}
We individually embed the components of an email, which are the username, the subdomain and the topdomain. 
The embeddings of the sub- and topdomains are generated with one hot encoding of the 60\% most common domains. 
The remaining domains are evenly distributed among an extra $20$ buckets by hashing. 
This reduces the impact of rare domain names and grants flexibility to the model when facing unseen domains.

For the username, we use both a semantic embedding and a character-level embedding. The semantic embedding is based on the fasttext model \cite{fasttext}. 
This allows the model to use hidden semantic information learned by the pretrained model in a username that aids to shape the conditional password space of a user.

The character-level encoder is a two-layer LSTM with $384$ units followed by a dense layer. 
It mimics the decoder architecture in its tokenization of the characters, using one hot encoding over the alphabet of the dataset.

\subsubsection{Decoder}
The embedded email components together with a positional encoding are passed to a self-attention layer, which completes the encoder part of the attention model. 
We use positional encoding to differentiate the source of each embedding to the attention layer.
The output of the encoder is passed through a dense layer and forms the initial state of a two-layer LSTM decoder with the same specification as the base model.
See Appendix A. for more details.

\subsection{Training setup}
We split the data set into train, validation, and test data in magnitudes of $202:1:186$ million. 
For the training, we employ the Adam optimizer \cite{kingma2017adam} and the learning rate is scheduled to decrease after every $2$ million steps of training from $1e^{-3}$ to $1e^{-4}$ to $5e^{-5}$ and finally to $3e^{-5}$.
The model stops the training when the validation loss did not decrease for more than $30$ iterations. 

\subsection{Inference}
\label{subsec:Inference}

\subsubsection{Calculating probabilities}
To calculate the probabilities of a character encoded password $p = [c_1, c_2, c_3, ..., c_n]$ under the base model $\mathcal{M}_{b}$, we obtain the probability distribution for each character under $\mathcal{M}_{b}$ and calculate the password probability as the conditional probability of each character given the previous characters: 

\begin{equation}
\begin{aligned}
\mathcal{P}_{\mathcal{M}_{b}}(p) & = \mathcal{P}_{\mathcal{M}_{b}}([c_1, c_2, c_3, ..., c_n])  \\
                                 & =\mathcal{P}_{\mathcal{M}_{b}}(c_1)                          \cdot \mathcal{P}_{\mathcal{M}_{b}}(c_2|c_1) \\
                                &\quad \cdot ... \cdot \mathcal{P}_{\mathcal{M}_{b}}(c_n|c_{n-1},..., c_1)  \\
\end{aligned}
\end{equation}

For the context-sensitive model  $\mathcal{M}_{c-s}$, we additionally condition the probability on a given character encoded email $e$: 
\begin{equation}
\begin{aligned}
\mathcal{P}_{\mathcal{M}_{c-s}}(p|e) &  = \mathcal{P}_{\mathcal{M}_{c-s}}([c_1, c_2, c_3, ..., c_n]|e) \\
                                         & = \mathcal{P}_{\mathcal{M}_{c-s}}(c_1|e) \cdot \mathcal{P}_{\mathcal{M}_{c-s}}(c_2|c_1, e) \\
                                         & \quad \cdot ... \cdot \mathcal{P}_{\mathcal{M}_{c-s}}(c_n|c_{n-1},..., c_1, e) \\
\end{aligned}
\end{equation}

\subsubsection{Compare probabilities}
\label{subsec:Inference_log_buckets}
To compare the two models, we first look at the different probabilities that they assign to a password.
We create \textit{logarithmically} equal spaced buckets ranging from $0$ to $1$ and sort each probability generated under the two models in its corresponding bucket.
To investigate which model assigns a higher probability to a password, we then compare the buckets rather than the raw probabilities. 
This offers robustness in determining the superiority of a given model since the logarithmic spacing accounts for a fairer assessment of the difference between each assigned probability in the regime of low as well as high probabilities.

Independent of the logarithmic buckets, we also introduce the metric of a difference score, calculated by the formula 
\begin{equation}
\mathcal{D}_{\text{c-s}, \text{base}}(x) = \biggl| \frac{\mathcal{P}_{\text{c-s}}(x) - \mathcal{P}_{\text{base}}(x) } {\mathcal{P}_{\text{c-s}}(x) + \mathcal{P}_{\text{base}}(x) } \biggr|, 
\end{equation}

which allows us to assess the relative difference between two predicted probabilities under the two models, respectively.


\subsubsection{Calculating guess numbers}
\label{subsec:guess_numbers}
In order to evaluate the password strength via simulating a guessing attack, we have to calculate the guess number of a password. 
A guess number denotes how many guesses it would take an attacker to try a given password under the assumption that all password guesses are arranged in descending order of their likelihood under a given model.
Therefore it serves as an indicator of the strength of a password and of the respective model.
Let $\Gamma$ denote the (possibly infinite) set of all allowed passwords.
Then the guess number \textbf{$\mathcal{G}$} for password $x$ under model  $\mathcal{M}$ is:

\begin{equation}
 \mathcal{G}_{\mathcal{M}}(x) = |\{y \in \Gamma: \mathcal{M}(y) > \mathcal{M}(x)\}|
\end{equation}

An efficient way to extract the guess number is by Monte Carlo simulations ~\cite{DellAmico2015}.
In a Monte Carlo simulation, a sample $S$ consisting of $n$ passwords is generated via ancestral sampling.  
For the base model $\mathcal{M}_{b}$, the guess number $\mathcal{G}$ of a password $x$ is then calculated as:
\begin{equation}
\begin{aligned}
 & \mathcal{G}_{\mathcal{M}_{b}}(x) = \\ 
  & \quad \mathlarger{\sum}_{x_i \in S}{
    \begin{cases}
      \Big( \mathcal{P}_{\mathcal{M}_{b}}(x_i) \cdot n \Big) ^{-1} & \text{$\mathcal{P}_{\mathcal{M}_{b}}(x_i) > \mathcal{P}_{\mathcal{M}_{b}}(x)$}\\
      0 & \text{otherwise}
    \end{cases} } 
    \end{aligned}
\end{equation}

Regarding the context-sensitive model $\mathcal{M}_{c-s}$, we have to condition the sampled passwords on a given email $e$:
\begin{equation}
\begin{aligned}
 & \mathcal{G}_{\mathcal{M}_{cs}}(x) = \\
 & \quad \mathlarger{\sum}_{x_i \in S}{
    \begin{cases}
      \Big( \mathcal{P}_{\mathcal{M}_{cs}}(x_i|e) \cdot n \Big) ^{-1} & \text{ $\mathcal{P}_{\mathcal{M}_{cs}}(x_i|e) > \mathcal{P}_{\mathcal{M}_{cs}}(x|e)$}\\
      0 & \text{otherwise}
    \end{cases} } 
    \end{aligned}
\end{equation}

If we employ the context-sensitive model, a new Monte Carlo sample of passwords has to be created for every different email address, making the procedure of testing multiple email addresses computationally quite expensive. 
Note that the Monte Carlo simulation is only possible due to the auto-regressive nature of the model, for example a GAN that generates an entire word and not a character at a time would not be able use this method.

\section{Results}
\label{section:results}
\subsection{Comparison base- and context-sensitive model}
An established metric to compare two language models is the perplexity score.
We calculate it by taking the exponent of the loss of a sample of passwords from the test set and obtain the mean and the median perplexity as comparative statistics.
Based on a randomly sampled subset of the test set of size $100,000$, we find that the mean perplexity score of the context-sensitive model is \textbf{12.785} and $12.97$ for the base model. 
The median is \textbf{9.62} and $9.81$, respectively.
The perplexity score being lower under the context-sensitive model is a first indicator for the superiority of this model.

\begin{figure}
    \includegraphics[width=\linewidth]{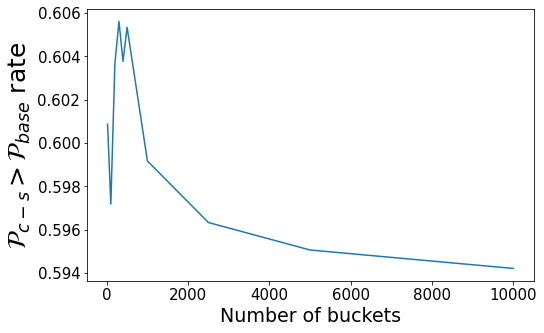}
    \caption{Change of the probability rate $\mathcal{P}_{c-s} > \mathcal{P}_{base}$ with varying number of buckets.}
    \label{fig:buckets_plot}
\end{figure}

Another indicator is given by comparing the two probabilities under both models for each test data pair, as described in Section~\ref{subsec:Inference_log_buckets}.
The bucket size being a potential hyperparameter, we investigate in Figure~\ref{fig:buckets_plot} how changing it influences the superiority rate of the context-sensitive model.
We can see that for a random sample of size $500,000$ from the test data, the context-sensitive model consistently assigned a higher probability to the passwords than the base model for all bucket sizes in $\mathcal{P}_{c-s} > \mathcal{P}_{base}  \approx \textbf{60\%}$ of the cases. 
Our results are therefore independent of any particular bucket size.

\begin{figure}
    \includegraphics[width=\linewidth]{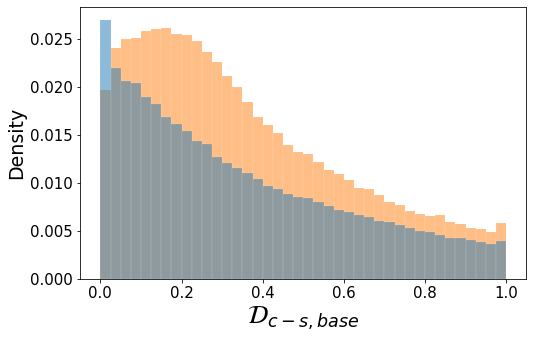}
    \caption{Distribution of the absolute difference scores $\mathcal{D}_{\text{c-s}, \text{base}}$ for a randomly sampled subset of the test data of size $500,000$ under the base and the context-sensitive model. 
    Blue: passwords that were assigned a higher probability under the base model. 
    Orange: passwords that were assigned a higher probability under the context-sensitive model. }
    \label{fig:difference_score}
\end{figure}

Looking at the probability differences score $\mathcal{D}_{\text{c-s}, \text{base}}$ calculated from the test data, we can observe in Figure~\ref{fig:difference_score} that the context-sensitive model (orange) surpasses the base model (blue) for all possible differences, except for a difference score of around $0.0$.
This shows that the predicted probability of a password is significantly higher under the context-sensitive model than under the base model.


\subsection{Guess numbers}

\begin{table*}[!t]
\caption{Probabilities $\mathcal{P}$ and guess numbers $\mathcal{G}$ for the $10$ samples from the test set with a difference score $\mathcal{D}_{\text{c-s}, \text{base}} > 0.99$.
Further note that the password leak is publicly available and therefore reporting the username-password pair does not represent a privacy violation.
}
\label{tab:guess_numbers}
\centering
\begin{tabular}{llccccc}
\toprule
$\texttt{Email}$ & $\texttt{Password}$ & $\mathcal{P}_{base}$ & $\mathcal{P}_{c-s}$ & $\mathcal{G}_{base}$ & $\mathcal{G}_{c-s}$ & $\log(\frac{\mathcal{G}_{base}}{\mathcal{G}_{c-s}})$ \\ 
\midrule
celticbiker80@gmail.com & \#E\$R\%T6y7u8i(O)P & 1e-40 & 1e-28 & 9e34 & 8e23 & 25 \\ 
lp.lp@lplppl.lp & LPLPLPLPLPLPLP & 1e-30 & 6e-21 & 6e25 & 2e17 & 20 \\ 
ai20092011@mail.ru & 00000qqqqq|Good & 4e-36 & 2e-28 & 7e30 & 1e24 & 16 \\ 
huangping20032@yahoo.com & HAPhgp520YINGyan & 2e-32 & 3e-21 & 3e27 & 7e21 & 13 \\ 
responsiblewoman@yahoo.com & RESPONSIBLE & 1e-32 & 1e-27 & 4e27 & 1e23 & 10 \\ 
urich37@yandex.ru & 37rus89644909462a & 1e-26 & 1e-21 & 1e22 & 1e18 & 10 \\ 
huliteryatsya@mail.ru & 15.10.1996.13.10.1996 & 1e-30 & 2e-25 & 7e25 & 3e21 & 10 \\ 
chocolate87@live.it & dolcecioccolata & 2e-17 & 1e-12 & 3e14 & 3e10 & 9 \\
famillerigaud@live.fr & famillerigaud & 9e-18 & 2e-13 & 6e14 & 1e11 & 8 \\ 
stereo@x-city.com.ua & stereostereostereo & 5e-21 & 4e-17 & 3e17 & 2e14 & 7 \\ 
\bottomrule
\end{tabular}
\vspace{1em}

\end{table*}

\normalsize

A popular method to compare the effectiveness of password attack models are guess numbers \cite{DellAmico2015, Melicher2016, Pal}. 
Computed as outlined in Section~\ref{subsec:guess_numbers}, we present in Table~\ref{tab:guess_numbers} the guess numbers under both models for the $10$ samples from the test data with a high difference score $\mathcal{D}_{\text{c-s}, \text{base}}$.
In the last column we compute the logarithm of the guess number under the base model $\mathcal{G}_{base}$ subtracted from the logarithm of the guess number under the context-sensitive model $\mathcal{G}_{c-s}$, i.e. $\log(\frac{\mathcal{G}_{base}}{\mathcal{G}_{c-s}}) = \log(\mathcal{G}_{base}) - \log(\mathcal{G}_{c-s})$. 
This number tells us how many orders of magnitude the guess number is higher under the base model than under the context-sensitive model. 
We can see that the context-sensitive model offers a considerable advantage when guessing the passwords of these vulnerable email addresses.

Note how the first displayed password $\texttt{\#E\$R\%T6y7u8i(O)P}$ could be evaluated as strong by a naive password meter, based on characteristics such as length, randomness, and special characters. 
However, given the email address, the password guess number decreases by $25$ orders of magnitude.
The email domain hints at the user being from the US. 
Indeed, by inspecting a standard US-QWERTY-based keyboard, we can observe that the password was created by iteratively selecting characters from the number and highest character row, while pressing the SHIFT key at the beginning and the end.
This explains the detected decrease in the password's strength in terms of required guesses by the context-sensitive model compared to the base model.

Under the viewpoint of length, the second password, $\texttt{LPLPLPLPLPLPLP}$ can be evaluated as strong. Given that we know the username, topdomain, and subdomain contain the substring $\texttt{lp}$, however, the password's guess number decreases by $20$ orders of magnitude.
Another pattern we observe by looking at the email $\texttt{huangping20032@yahoo.com}$ is that the corresponding password $\texttt{HAPhgp520YINGyan}$ is mostly composed of the letters and digits of the username.
Lastly, knowing the user's topdomain hints at the password containing words in the specific target language.
For the email address $\texttt{chocolate87@live.it}$ the corresponding password $\texttt{dolcecioccolata}$ is in Italian, as indicated by the topdomain $it$.

By investigating the $30$ samples from the test data with the highest difference score (c.f. Appendix Table~\ref{table:better_comp_ctx} for the full table), we can state more generally that knowing the email address of the target decreases the guess number on average by $11.5$ orders of magnitude compared to the base model.
The median decrease in magnitude is $10$.
When looking at the username-password pairs we find that $20\%$ of the users are reusing parts of their username in the password, where $7\%$ use their entire username concatenated with itself as their password.
Furthermore, we can see that $67\%$ of the users use meaningful sentences, phrases, or the same word concatenated with itself as their password.

Our results demonstrate that context aware models can exploit the fact that reusing a user's username, parts of their username, or words in a specific target language as indicated by the topdomain of the email weakens the strength of their passwords by several orders of magnitude.



\subsection{When the base model performs better}

The user email can confuse the context aware model in two main cases: if the user uses random characters and if the user chooses a very common password such as $123456$. 
In the first case both the context-aware and base model show similar performances, c.f. Table~\ref{table:worst_base} and Table~\ref{table:worst_ctx}, because both models cannot guess a random sequence of characters. 
The second case reflects the main difference between the two models. 
The base model learns the general distribution of passwords, which will output the most common password in the dataset with highest probability, c.f. Table~\ref{table:best_base}, while the context-aware model prioritizes groups of users with similar usernames that generate similar password, c.f. Table~\ref{table:best_ctx}).

If we look at the pair where the probability difference of the base model is much higher than the context-aware one (c.f. Table~\ref{table:better_comp_base}) we can spot short pseudo-random passwords such as $d9Zufqd92N$ which is shared by at least 4 emails:
\begin{itemize}
    \item $daleek13@lyme.in$
    \item $francisqg20@kamryn.tia.inxes.in$
    \item $biancabt34@mail.oplog.in$ 
    \item $junevb16@chi.blognet.in$
\end{itemize}
Note that all usernames have the same construction: $name$-$2$ $random$ $characters$-$2$ $digits$. This indicates that the users have been generated automatically and that fake profiles can be spotted looking at the difference of the evaluations of both models.

\subsection{Relevance of the contextual input}

\subsubsection{Average attention weights highlight username importance}
The attention weight outputs for each head averaged over $120,000$ random samples from the test set as visualized in Table~\ref{table:attention_score} provide inside into the inner workings of the context-sensitive model. 

\begin{table}[]
\caption{We show the cumulative attention score for each part of the email. The cumulative attention score is averaged over $64,000$ random samples from the test set. }
\label{table:attention_score}
\centering

\begin{tabular}{lllll}
\toprule
\multicolumn{1}{c}{} &
  \multicolumn{1}{c}{\begin{tabular}[c]{@{}c@{}}username\\ LSTM\end{tabular}} &
  \multicolumn{1}{c}{\begin{tabular}[c]{@{}c@{}}sub\\ domain\end{tabular}} &
  \multicolumn{1}{c}{\begin{tabular}[c]{@{}c@{}}top\\ domain\end{tabular}} &
  \multicolumn{1}{c}{\begin{tabular}[c]{@{}c@{}}username\\ Fasttext\end{tabular}} \\ 
  \midrule
\multicolumn{1}{c}{\begin{tabular}[c]{@{}c@{}}cumulative\\ attention\\score\end{tabular}} &
  \multicolumn{1}{c}{\textbf{3.41}} &
  \multicolumn{1}{c}{2.64} &
  \multicolumn{1}{c}{2.91} &
  \multicolumn{1}{c}{3.27} \\  
  \bottomrule
\end{tabular}

\end{table}

These findings highlight the username's importance in shaping the latent space of the password distribution given a specific username.
This implies that usernames and passwords are far from being independent of each other, but that the model was able to extract a great amount of structure from the username.
The attention weights themselves do not tell us what specific structure that was, however, one explanation we found in an exploratory analysis of the data is that users tend to reuse parts of their username or repetitive patterns in their password.

The attention weights also revealed that the model was able to condition the output density on the sub- and topdomain.
A possible explanation can be found in the domains revealing information about the nationality of the user since users tend to use email providers from their respective country of origin.
This speculation can be further confirmed by an analysis of the hidden states of the decoder, as we will see in the following section.

\subsubsection{Sub- and topdomain influence}
Another way to understand the structure that the context-sensitive model has learned from the additional data is to inspect the hidden states of the its decoder, which we visualize with the TSNE algorithm in Figure~\ref{fig:subdomain_hidden} and Figure~\ref{fig:topdomain_hidden} \cite{vanDerMaaten2008}.
The model was able to identify clusters for the sub- as well as the topdomain.
Sometimes, as in the case of the topdomain $\texttt{"com"}$ and $\texttt{"ru"}$, there are multiple separate clusters for the same domain.
This implies that the model was able to identify different regularities within a domain that facilitate its task to predict the probability of a certain password.

\begin{figure}
    \centering
    \includegraphics[scale=2.0, page=4] {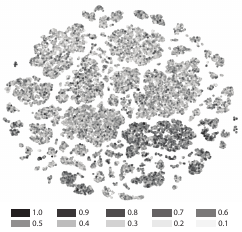}
    \caption{$P_{c-s} > P_{base}$. Difference score $\mathcal{D}_{\text{c-s}, \text{base}}$ for the test data entries where the relation $P_{c-s} > P_{base}$ holds.
    We can observe that there are clusters for which the context-sensitive model assigns a much higher probability to the passwords than the base model.}
    \label{fig:difference_A_bigger_B}
\end{figure}

Investigating this claim, we colored the same test data points according to the difference in assigned probability $\mathcal{D}_{\text{c-s}, \text{base}}$ under the two models and display the results in Figure~\ref{fig:difference_A_bigger_B}.
We can see that there are clusters for which the context-sensitive model assigns a much higher probability to the passwords than the base model, meaning that these classes are especially vulnerable to a context-sensitive password guessing attack.

\begin{figure*}
    \centering
    \includegraphics[scale=0.4]{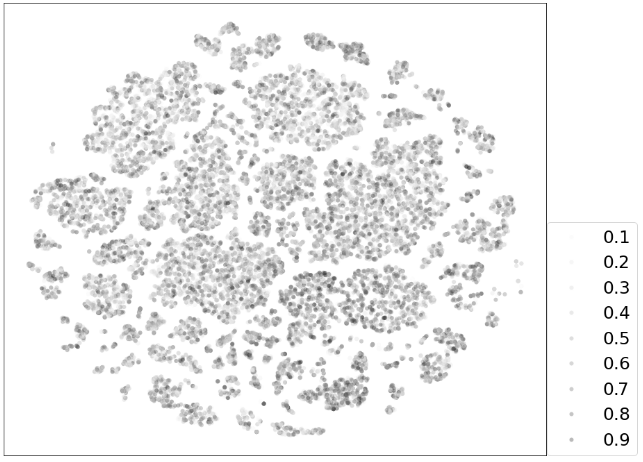}
    \caption{$P_{c-s} < P_{base}$. Difference score $\mathcal{D}_{\text{c-s}, \text{base}}$ for test data entries where the relation $P_{c-s} < P_{base}$ holds.
    There is no clear cluster for which the base model assigns a higher probability to the passwords than the context-sensitive model}
    \label{fig:difference_B_bigger_A}
\end{figure*}

To investigate what these clusters have in common, we can refer back to Figure~\ref{fig:subdomain_hidden} and Figure~\ref{fig:topdomain_hidden}. 
The clusters with a high difference in predicted probability have either the subdomain $\texttt{"mail"}$ mixed with $\texttt{"gmail"}$, or the subdomain $\texttt{"rambler"}$ or $\texttt{"aol"}$.
Interestingly, there are clusters with subdomain $\texttt{"yahoo"}$ or $\texttt{"hotmail"}$ that have a high probability difference but are separated from other clusters with that subdomain.
Regarding the topdomain we observe that there are high difference clusters with either the topdomain $\texttt{"ru"}$, $\texttt{"fr"}$, $\texttt{"cn"}$, or $\texttt{"com"}$.
In contrast to that, there are no visibly identifiable clusters for the case that the base model assigned a higher probability to a test data point than the context-sensitive model, c.f. Figure~\ref{fig:difference_B_bigger_A}.

These findings indicate that the context-sensitive model was able to extract structure from the available email. 
It could even identify subclusters within different domains for which it learned regularities in the email address that allowed it to expect the corresponding password with a much higher probability than the base model.
This is another compelling result highlighting how much structure there is to be found in public information that largely diminish the strength of a target password.

\section{Conclusion}

We were able to find compelling evidence in favor of the hypothesis that additional information weakens the strength of a user's password.
%
%
The context-sensitive model learned regularities in the email address that allowed it to expect a password with a much higher probability than the base model.
The context-sensitive model can identify clusters of users whose passwords are easy to guess, making those groups extremely vulnerable to password guessing attacks with a context-aware model.
\par
A limitation of our analysis is faced by the presence of password managers that allow the user to use truly randomly generated passwords. 
However, there still needs to be one master password that is to be chosen and remembered by the user, and recent password manager breaches underpin the usefulness of our analysis.

Conclusively, this shows that password models that are currently not exploiting public and easily retrievable information about users, such as their email addresses, overestimate the real strength of a password and fail to model realistic password attacks.


With our research, we intend to highlight the importance of deep learning in simulating attacks that exploit data leaks and encourage the development of more sophisticated password models that take into account a wide range of contextual information.





\newpage

\balance
\bibliographystyle{plain}
\bibliography{references}

\appendices

\section{Appendix}

\subsection{Training setup}
We chose the generous train:validation:test split of $202:1:186$ million due to the abundance of available data and to ensure that the obtained test score for our method is representative.

\subsection{Base model}

\begin{figure}[h]
  \centering
    \includegraphics[scale=1.0, page=1] {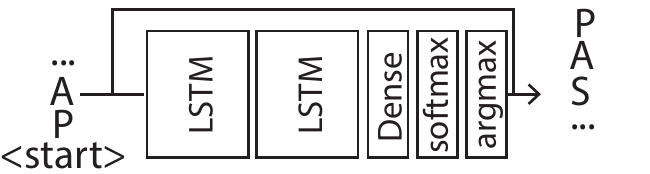}
    \caption{Model architecture of the base model. 
    We chose to follow the architecture proposed by Melicher et al. \cite{Melicher2016}, since their work represents an important milestone in the password security research.
    The $2$ LSTM layers have $512$ units each. 
    The output size of the dense layer corresponds to the length of our vocabulary which is the number of ASCII characters. 
    The last softmax and argmax layers are used to select the next character with the highest probability to take.}
    \label{fig:basemodel}
\end{figure}

The base model is not context-aware. 
It is an auto-regressive model, that generates one character at a time based on the previously generated character. 
See Figure \ref{fig:basemodel} for a detailed description of the model architecture. 

\subsection{Context aware model}
\subsubsection{Preprocessing}

We remove any emails containing non-ASCII characters and with a username that is longer than $32$ characters.
We also remove duplicate entries and email addresses that appear more than $100,000$ times in the database. 
Since we split the email address into its components based on the "@" symbol, we filter emails that contain more than one of those. Similarly, we filter emails and passwords with the ":" symbol as it is the delimiter used in the dataset to split email and passwords.
After the preprocessing, we are left with $390,147,060$ email-password pairs, where each data entry has the format $\texttt{username@subdomain.topdomain:password}$.

In our context-sensitive attention model, we use two different encodings for the username: a character-level one and a word embedding based on the multi-lingual fasttext model developed at Facebook \cite{bojanowski2016enriching}.
We rely on a multi-lingual embedding since the password leak we use as our data origin from a multitude of different countries.
In the preprocessing of the username for the fasttext model, we split the username based on any non-alphabetic token and remove all those non-alphabetic tokens. 
For instance, the username \textit{john.doe} would be split into \textit{[john, doe]} and the username \textit{spaceexplorer123456} would be processed as \textit{[spaceexplorer]}.

\section{Results}

\subsection{Attention score matrix}

The average attention score of the context aware model in Table  \ref{table:attention_score} show that each head is focusing on embeddings originating from the username part (LSTM and fasttext) compared to the sub abd top domain.

\begin{table}[]
    \caption{ email-password pairs with highest probability with the $base$ model }
    \label{table:best_base}
    \vspace{.1cm}
    \centering
    \resizebox{7.7cm}{!}{
    \begin{tabular}{llll}
     \toprule
    email                            & $pwd$ & $\mathcal{P}_{ctx}$ & $\mathcal{P}_{base}$  \\  
    \midrule
    kara\_kartal541@hotmail.com	     & 12345     &  \textbf{0.018454}	& 0.018093  \\  
    travist\_yudh@yahoo.com	         & 12345     &  \textbf{0.018592}	& 0.018093 \\  
    loveyougodpidra@yahoo.com	     & 12345     &  \textbf{0.020903}	& 0.018093 \\  
    jesz1@tlen.pl                    & 12345     &  0.015960	& \textbf{0.018093} \\  
    vnmuta5@gmail.com                & 12345     &  \textbf{0.029174}	& 0.018093 \\  
    niko94@live.it                   & 12345     &  \textbf{0.022636}	& 0.018093  \\  
    rgd2@yandex.ru                   & 12345     &  \textbf{0.024919}	& 0.018093  \\  
    almetadag\_9643@yandex.ru         & 12345     &  \textbf{0.029047}	& 0.018093  \\  
    berence\_08@hotmail.com           & 12345     &  \textbf{0.018241}	& 0.018093  \\  
    hiphoper@rambler.ru              & 12345     &  \textbf{0.026887}	& 0.018093  \\  
    \bottomrule
    \end{tabular}}
\end{table}

\begin{table}[]
    \caption{ email-password pairs with highest probability with the $ctx$ model}
    \label{table:best_ctx}
    \centering
    \resizebox{7.5cm}{!}{
    \begin{tabular}{llll}
    \toprule
    email                            & $pwd$ & $\mathcal{P}_{ctx}$ & $\mathcal{P}_{base}$\\ 
    \midrule
    wangzi160@tianya.cn          & 123456     &  \textbf{0.087573}	& 0.01577	 \\ 
    qingtaowang2003@tianya.cn    & 123456     &  \textbf{0.085443}	& 0.01577	  \\ 
    sunguirong1234@tianya.cn     & 123456     &  \textbf{0.085423}	& 0.01577	  \\ 
    vbnm567@tianya.cn            & 123456     &  \textbf{0.082656}	& 0.01577  \\ 
    liuyulun123@tianya.cn        & 123456     &  \textbf{0.082291}	& 0.01577	 \\ 
    yokoyico110@tianya.cn        & 123456     &  \textbf{0.081175}	& 0.01577	 \\ 
    akira123@tianya.cn           & 123456     &  \textbf{0.079154}	& 0.01577	 \\ 
    mtwj02@tianya.cn             & 123456     &  \textbf{0.066099}	& 0.01577 \\ 
    sailormoon1@tianya.cn        & 123456     &  \textbf{0.065521}	& 0.01577	\\ 
    ybyq86@tianya.cn             & 123456     &  \textbf{0.063941}	& 0.01577  \\ 
    \bottomrule
    \end{tabular}}

\end{table}

\begin{table*}[t]
    \caption{ email-password pairs with the lowest probability under the $ctx$ model}
    \label{table:worst_ctx}
    \centering
    \resizebox{15cm}{!}{
    \begin{tabular}{llll}
    \toprule
    email                            & $pwd$ & $\mathcal{P}_{ctx}$ & $\mathcal{P}_{base}$  \\ 
    \midrule
    it@capnajax.com                           & iB]Wzx;V3u3\%cw>w)<Mz2=3QPJW479      &  2.95e-78	& \textbf{2.92e-75}  \\ 
    yrm2e-lkgjq-2x87v-2idrl-3myeq@live.com    & YRM2E-LKGJQ-2X87V-2IDRL-3MYEQ        &  \textbf{7.79e-59}	& 1.18e-59 \\ 
    www.frank\_ungemach@web.de                & bc?watfngjl\&-wvs-qmv-zv\$v	         &  1.64e-56	& \textbf{3.03e-52} \\ 
    whiskas-forever2008@yandex.ru             & F8910fzZkvirus005Lerik@vIel          &  1.68e-54	& \textbf{1.51e-53}  \\ 
    w@razryv.1gb.ru                           & w@razryv.1gb.ruw@razryv.1gb.ru       &  2.07e-54	& \textbf{1.73e-55} \\ 
    peri.odi.c.al.we.r.y@yahoo.com            & ox582HzxsEE6b1iiGNevGJzLfd           &  \textbf{3.24e-53}	& 1.29e-53  \\ 
    bebek\_citra@yahoo.com                     & q\$Lnpo623X67j7HxeJ\$vb6u            &  8.57e-52	& \textbf{2.16e-51}  \\ 
    lukinskaya.diana@yandex.ru                & jAmBvAVd8V5S6LVSnhDhFhFgOM           &  \textbf{1.53e-51}	& 8.20e-53  \\ 
    qwuy23ioy3qtuwgetjhgewhj@mail.ru          & qwkjhr32uirio132uriqwhrthewj         &  2.05e-49	& \textbf{6.15e-48}  \\ 
    tolstenkova@mail.ru                       & alge6ra1p0chat0kanal13u94942443	     &  \textbf{4.63e-49}	& 9.46e-53  \\ 
    \bottomrule
    \end{tabular}}
 \end{table*} 
 
 \begin{table*}[t]
     \caption{ email-password pairs with the lowest probability under the $base$ model }
    \label{table:worst_base}
    \centering
    \resizebox{15cm}{!}{
    \begin{tabular}{llll}
    \toprule
    email                            & $pwd$ & $\mathcal{P}_{ctx}$ & $\mathcal{P}_{base}$ \\ 
    \midrule
    it@capnajax.com                  & iB]Wzx;V3u3\%cw>w)<Mz2=3QPJW479        &  2.95e-78	& \textbf{2.92e-75}  \\ 
    yrm2e-lkgjq-2x87v-2idrl-3myeq@live.com  & YRM2E-LKGJQ-2X87V-2IDRL-3MYEQ	  &  \textbf{7.79e-59}	& 1.18e-59  \\ 
    w@razryv.1gb.ru                  & w@razryv.1gb.ruw@razryv.1gb.ru         &  \textbf{2.07e-54}	& 1.73e-55  \\ 
    peri.odi.c.al.we.r.y@yahoo.com   & ox582HzxsEE6b1iiGNevGJzLfd	          &  \textbf{3.24e-53}	& 1.29e-53  \\ 
    whiskas-forever2008@yandex.ru    & F8910fzZkvirus005Lerik@vIel            &  1.68e-54	& \textbf{1.51e-53}  \\ 
    lukinskaya.diana@yandex.ru       & jAmBvAVd8V5S6LVSnhDhFhFgOM             &  \textbf{1.53e-51}	& 8.20e-53  \\ 
    tolstenkova@mail.ru              & alge6ra1p0chat0kanal13u94942443	      &  \textbf{4.63e-49}	& 9.46e-53  \\ 
    www.frank\_ungemach@web.de        & bc?watfngjl\&-wvs-qmv-zv\$v             &  1.64e-56	& \textbf{3.03e-52}  \\ 
    bebek\_citra@yahoo.com            & q\$Lnpo623X67j7HxeJ\$vb6u	          &  8.57e-52	& \textbf{2.16e-51}  \\ 
    cleveland515@yahoo.com           & 1nn0c3nttVmwxhcRHsZwhsaY	              &  \textbf{3.72e-46}	& 1.36e-48  \\ 
    \bottomrule
    \end{tabular}}
\end{table*}

\begin{table*}[]
\caption{  selected email-password pair and their respective guess number and probability score, where the probability of $ctx$ is better than $base$ }
\label{table:better_comp_ctx}
\centering
\resizebox{13cm}{!}{
\begin{tabular}{llllll}
\toprule
email                            & $pwd$ & $\mathcal{P}_{ctx}$ & $\mathcal{P}_{base}$ & $\mathcal{G}_{ctx}$ & $\mathcal{G}_{base}$ \\ 
\midrule
celticbiker80@gmail.com          & \#E\$R\%T6y7u8i(O)P	      &  \textbf{1.62e-28}	& 1.20e-40 & 	\textbf{8.18e+23} &	9.33e+34 \\  
lp.lp@lplppl.lp                  & LPLPLPLPLPLPLPLPLPLP 	& \textbf{6.71e-21} & 	1.15e-30 & \textbf{2.21e+17} &	6.56e+25 \\  
mikeoogod@gmail.com               & AZSXdcfv~!@            &	\textbf{7.04e-18}	& 7.54e-27 & \textbf{8.08e+14} &	2.74e+22 \\  
imant94@mail.ru & yfever[bvelhj & \textbf{1.37e-19} & 2.97e-27 & \textbf{2.77e+16} & 5.73e+22 \\  
huangping20032004@yahoo.com.cn & HAPhgp520YINGyan & \textbf{5.16e-26} & 2.13e-32 & \textbf{6.85e+21} & 3.24e+27 \\  
garik1@rambler.ru & igadministrator &	\textbf{6.91e-13} & 1.01e-18 & \textbf{3.87e+10} & 3.86e+15 \\  
wifnqiw678@aol.com & 8ix6S1fceH	& \textbf{3.05e-09} & 	7.46e-15 & \textbf{2.02e+07} & 1.87e+12 \\  
etramdirai@gmail.com & DVGGhejAxMQSfQc4 & \textbf{6.21e-30} & 1.58e-35 & \textbf{1.72e+25} & 1.45e+30 \\  
ilve\_\_34@yahoo.de & h2eternity & \textbf{3.25e-10} & 1.16e-15 & \textbf{1.55e+08}  & 9.30e+12 \\  
huliteryatsya@mail.ru  & 15.10.1996.13.10.1996.88.R. & \textbf{2.29e-25} & 1.13e-30 & \textbf{3.18e+21} & 6.56e+25 \\  
kiki\_duarte\_2@web.de & dothepeanutbuttajelly & \textbf{1.04e-19} & 5.49e-25 & \textbf{3.43e+16} & 7.10e+20 \\  
emanuel-sgg@yahoo.de & xx\_candyflavored\_xx & \textbf{1.41e-18} & 8.66e-24 & \textbf{2.95e+15} & 6.40e+19 \\  
michelineb@lycos.de & babyxtazxinxbabyxbluex & \textbf{1.38e-28} & 8.64e-34 & \textbf{5.78e+23} & 4.86e+28 \\  
urich37@yandex.ru & 37rus89644909462a & \textbf{1.28e-21} & 1.25e-26	& \textbf{9.72e+17} & 1.42e+22 \\  
carmine.nastro.gosp@web.de & takaministryofinfor& \textbf{7.77e-23} & 7.68e-28 & \textbf{1.26e+19} & 2.25e+23 \\  
donkacosta@hotmail.com & electrocolombo53 & \textbf{1.33e-14} & 1.69e-19 & \textbf{9.09e+08} & 1.59e+16 \\  
responsiblewomanfortoday@yahoo.com & RESPONSIBLE & 	\textbf{1.07e-27}  & 1.40e-32 & 	\textbf{1.30e+23}  & 3.75e+27 \\  
stefanny777@list.ru & 7sagittarius12 &	\textbf{4.06e-14} & 6.72e-19 & \textbf{1.32e+09} & 5.19e+15 \\  
gatasalvaje3871@web.de & thefreepicktoncampaign & \textbf{4.88e-25} & 8.07e-30 & \textbf{1.11e+21} & 1.47e+25 \\  
chocolate87@live.it & dolcecioccolata & \textbf{1.12e-12} & 1.86e-17 & \textbf{2.61e+10} & 3.12e+14 \\  
famillerigaud@live.fr & famillerigaud & \textbf{1.59e-13} & 9.30e-18 & \textbf{1.31e+11} & 5.84e+14 \\  
solaceguddah@gmail.com & S14ORIGINAL & \textbf{3.56e-14} & 2.53e-18 & \textbf{4.46e+11} & 2.22e+15 \\  
fail\_apas@yandex.ru & afbkmqwe & \textbf{2.39e-10} & 1.99e-14 & \textbf{3.63e+08} & 7.74e+11 \\  
cerdanyola46@gmx.de & emilythecelebrity & \textbf{1.17e-14} & 1.02e-18 & \textbf{1.51e+12} & 3.84e+15 \\  
fominane@yandex.ru & YTAjvbyf & \textbf{1.13e-12} & 1.01e-16 & \textbf{2.57e+10} & 6.43e+13 \\  
mathculot@hotmail.com & labradorlabrador & \textbf{7.91e-13} & 7.26e-17 & \textbf{3.36e+10} & 8.85e+13 \\  
despoinakord@yahoo.com & NEVERGIVEUP & \textbf{1.12e-11} & 1.05e-15 & \textbf{3.20e+09} & 1.00e+13 \\  
martolina\_\_92@web.de & trevotheawesome & \textbf{1.31e-13} & 1.25e-17 & \textbf{1.57e+11} & 4.58e+14 \\  
stereo@x-city.com.ua & stereostereostereo & \textbf{3.85e-17} & 4.09e-21 & \textbf{1.95e+14} & 3.23e+17 \\  
\bottomrule
\end{tabular} }
\end{table*}

\begin{table*}[]
\caption{  selected email-password pair and their respective guess number and probability score, where the probability of $ctx$ is worst then $base$ }
\label{table:better_comp_base}
\centering
\resizebox{13cm}{!}{
\begin{tabular}{llllll}
 \toprule
email                            & $pwd$ & $\mathcal{P}_{ctx}$ & $\mathcal{P}_{base}$ & $\mathcal{G}_{ctx}$ & $\mathcal{G}_{base}$ \\  
\midrule
olegator\_87@inbox.ru         & X3LUym2MMJ     &  5.57e-17	& \textbf{4.61e-08} & 	1.11e+13 &	\textbf{1.02e+06} \\  
rrikimiki@rambler.ru         & qawsed         &  1.02e-13	& \textbf{5.76e-05} & 	1.91e+11 &	\textbf{2.33e+02} \\  
iseudoir941@inbox.ru         & 7ugd5hip2j     &  6.11e-15	& \textbf{3.67e-07} & 	2.27e+11 &	\textbf{1.36e+05} \\  
bestheyu@163.com             & NBvBB32fa9     &  5.03e-19	& \textbf{6.38e-12} &  6.32e+15 &	\textbf{5.29e+09} \\  
moses.mai@netiq.com          & maivtxLINKEDIN &  5.23e-26	& \textbf{4.36e-19} & 	5.80e+21 &	\textbf{7.35e+15} \\  
daleek13@lyme.in     & d9Zufqd92N     &  1.21e-14	& \textbf{1.14e-09} & 	1.22e+12 &	\textbf{4.67e+07} \\  
mepotts@ovis.net             & milky[way]     &  2.29e-25	& \textbf{2.13e-20} & 	1.49e+21 &	\textbf{7.56e+16} \\  
shuyikan@163.com             & 9ol.0p;/       &  1.14e-20   & \textbf{5.33e-15} & 	1.30e+17 &	\textbf{2.55e+11} \\  
kazzaanna@yahoo.com          & djpleiyfzlshf  &  8.54e-24   & \textbf{2.88e-18} & 	6.40e+19 &	\textbf{2.04e+14} \\  
030811306@bk.ru              & yuzhaofang     &  1.03e-15   & \textbf{3.41e-10} & 	1.02e+13 &	\textbf{1.71e+08} \\  
mamotenko.arina@yahoo.com	 & vscxfcnkbds112296    & 2.40e-23  & \textbf{5.58e-18} & 	2.73e+19 &	\textbf{9.90e+14 } \\  
ponfilovitch@yahoo.com      & rty789fgh456vbn123xc & 1.56e-30  & \textbf{3.61e-25}      & 	3.44e+25 &	\textbf{1.08e+21} \\  
temka89228628680@rambler.ru  & fujitsusiemens & 1.18e-16    & \textbf{1.91e-11} & 	5.67e+13 &	\textbf{2.08e+08} \\  
francisqg20@kamryn.tia.inxes & d9Zufqd92N  & 7.36e-15    & \textbf{1.14e-09} & 	1.90e+12 &	\textbf{4.67e+07} \\  
cat2k2@msn.com	             & 7008J05A6AA47C68B    & 1.19e-29  & \textbf{1.49e-24}      & 	1.35e+25 &	\textbf{3.21e+20} \\  
endemoniado9@yahoo.fr        & testerqQ1!wW2"       & 5.94e-34  & \textbf{6.22e-29}      & 	6.17e+28 &	\textbf{2.90e+24} \\  
aparickmfrancesy@ymail.org   & pk3x7w9W	            & 2.02e-11  & \textbf{2.06e-06}      & 	2.00e+09 &	\textbf{1.75e+04} \\  
daleek13@lyme.minemail.in     & d9Zufqd92N           & 1.21e-14  & \textbf{1.14e-09}      & 	1.22e+12 &	\textbf{4.67e+07} \\  
htaz@pengelan123.com         & pk3x7w9W             & 2.30e-11  & \textbf{2.06e-06}      & 	1.79e+09 &	\textbf{1.75e+04} \\  
biancabt34@mail.oplog.in & d9Zufqd92N       & 7.34e-15  & \textbf{1.14e-09}      & 	1.91e+12 &	\textbf{4.67e+06} \\  
lfibadjlank@gmail.com        & qdujvyG5sxa          & 3.66e-13  & \textbf{2.07e-06}      & 	6.56e+10 &	\textbf{1.73e+04} \\  
ncefehmdaahi@yahoo.com       & qdujvyG5sxa          & 3.92e-13  & \textbf{2.07e-06}      & 	6.21e+10 & \textbf{1.73e+04}\\  
q1ae0kdxwthrtoq@mail.ru      & X3LUym2MMJ           & 6.64e-17  & \textbf{4.61e-08}      & 	9.47e+13 &	\textbf{1.02e+06} \\  
perlet7@yahoo.com	         & intereforjoet@sbcglobal.net  & 2.41e-29  & \textbf{1.56e-23}  & 	8.32e+24 & \textbf{3.79e+19}\\  
g.l.a.n.du.l.arnri.e@swbell.net & d2xyw89sxj  & 5.46e-16    & \textbf{3.60e-11} & 	1.67e+12 &	\textbf{1.27e+09} \\  
unfoua@gmail.com             & djcrhtctyrf    & 1.33e-15    & \textbf{7.49e-11} & 	8.58e+12 &	\textbf{7.47e+08} \\  
junevb16@imp.chi.blognet.in  & d9Zufqd92N     & 2.00e-14    & \textbf{1.14e-09} & 	7.72e+11 &	\textbf{4.67e+07} \\  
juliasunshienvvpc@gmail.com  & okhueleigbe5   & 4.87e-22    & \textbf{2.39e-17} & 	2.45e+18 &	\textbf{2.38e+13} \\  
tcdz83@mail-s01.pl           & pk3x7w9W       & 4.23e-11    & \textbf{2.06e-06} & 	1.12e+09 &	\textbf{1.75e+04} \\  
\bottomrule
\end{tabular}}
\end{table*}

 For three out of the four heads the sub- and topdomains ($\texttt{SUB}$, $\texttt{TOP}$) as well as the fasttext embedding of the username ($\texttt{USER-EMB}$) pay attention to the character-embedded username ($\texttt{USER-CHAR}$).
 In the remaining head, all layers except for the fasttext embedding of the username attend to the fasttext embedding.

\begin{figure*}
     \centering
    \includegraphics[scale=1.2, page=3] {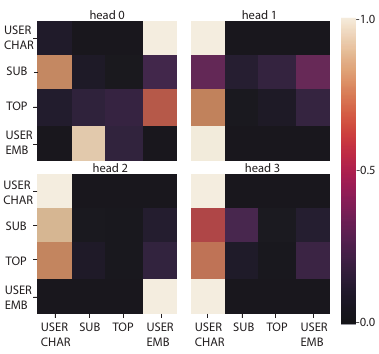}
    \caption{Average attention weights over $120,000$ random samples from the test set for each of the four heads.
        For three out of the four heads, all parts pay attention to the character embedded username ($\texttt{USER-CHAR}$).
        One head focuses on the username embedding generated by the fasttext model ($\texttt{USER-EMB}$).
        Minor attention is paid by three out of the four heads to the sub- and topdomain ($\texttt{SUB}$, $\texttt{TOP}$).}
    \label{fig:attention_weights}
\end{figure*}

\end{document}